\documentstyle[aps,prl,multicol]{revtex}
\include{psfig} 

\begin{document}

\draft \title{Test of the semischematic model for a liquid of linear 
molecules}
\author{Linda Fabbian$^{(1)}$, Rolf Schilling$^{(2)}$, Francesco Sciortino$^{(1)}$, 
Piero Tartaglia$^{(1)}$ and Christoph Theis$^{(2)}$}

\address{$(1)$ Dipartimento di Fisica and Istituto Nazionale
 per la Fisica della Materia, Universit\'a di Roma {\it La Sapienza},
 P.le Aldo Moro 2, I-00185 Roma, Italy}
\address{$(2)$ Institut f\"ur Physik, Johannes Gutenberg--Universit\"at, 
Staudinger Weg 7, D-55099 Mainz, Germany}

\date{\today}
\maketitle
\begin{abstract}
\noindent

We apply  to  a liquid of linear molecules
the semischematic mode-coupling model, previously introduced
to describe the center of mass (COM) slow dynamics of a network-forming 
molecular liquid. We compare the
theoretical predictions and numerical
 results from a molecular  dynamics simulation, both for the time  and
the wave-vector dependence of the COM 
density-density correlation function.
We discuss the relationship between the presented analysis and the 
results from an approximate  solution of the equations from molecular 
mode-coupling theory   
[R. Schilling and T. Scheidsteger,{\em Phys. Rev. E} {\bf 56} 2932 (1997)]. 
\end{abstract}


\begin{multicols}{2}

\section{introduction}
\label{introduction}
The mode-coupling theory (MCT)\cite{review-glass,gotze} has opened 
new perspectives in the theoretical understanding of the dynamic slowing
down characteristic of  supercooled glass-forming liquids\cite{review}.
MCT, originally developed to describe the structural relaxation 
in simple liquids, i.e. in liquids composed by particles interacting 
via spherically symmetric intermolecular
potentials, also offers a coherent picture of the slow 
dynamics in molecular glass forming liquids,  composed 
of asymmetric molecules.
The ability of MCT to model the onset of slow dynamics
in molecular liquids has prompted the need to extend the theory
to fully take into account the angular degrees of freedom.
The extension of MCT to a solute linear molecule 
in a solvent of spherical particles\cite{franosch-dumb} and to
molecular liquids\cite{schilling,theis}, which we refer to in
the following as molecular MCT (MMCT), has been recently
achieved.  The center of mass (COM) density-density
correlation function, which in MCT is the only relevant correlation
function, in MMCT becomes coupled to an
infinite hierarchy of rotational correlation functions, arising from
the expansion of the angular degrees of freedom in spherical harmonics.
The MMCT equations for COM and angular correlators have been solved
until now for systems of linear molecules under 
specific approximations\cite{schilling,theis,schilling-vigo}. Work is 
currently underway to improve the approximations for dumbbells 
and to calculate a  solution for the general case 
of molecules of arbitrary shape.
 
Recently, some of us proposed a parameterization of the role of the
rotational degrees of freedom and their effective 
coupling with the COM density\cite{ssm}.  This approach, which provides a
solvable set of coupled equations for the slow dynamics of the COM
density-density correlation function, has been named {\it
semischematic} because it retains all the $q$--dependence of the COM
correlators but condenses the coupling between COM and angular
correlation functions into a single $q$-independent parameter $\chi_R$.
A  detailed comparison between
the theoretical predictions of the model and results from a 
molecular dynamics (MD) simulation
has been performed for a network forming liquid\cite{ssm}, finding
an excellent agreement up to a cut-off $q$-vector value where the microscopic
geometric details become dominant.

In the long term, 
the exact solutions of the MMCT equations are to be preferred
because they do not require an {\it ad hoc}
estimate of the translational-rotational coupling and predict also 
the behavior of  the angular correlators. On the other hand,
in the short term the semischematic equations are rather appealing because of
their simplicity, for the minimal amount of input information required 
and because the complete time dependent solution  can be achieved with
present day computational facilities. 
Also, once the ability of the
semischematic model to describe the time evolution of the COM
correlation functions has been assessed by detailed comparison with
MD simulations, comparison
with experimental data is foreseeable, again due to the limited need
for structural input.  
For this reason, in this Article we solve
the semischematic model for a system of linear dumbbells
interacting through a Lennard-Jones (LJ) potential and compare the
predictions of the model with the corresponding quantities evaluated from
long MD simulations of the same liquid\cite{kob-dumb}.  The choice of
a liquid of dumbbells for which the MMCT equations have
been previously solved approximately \cite{schilling-vigo} allows at 
the same time a comparative discussion of the two theoretical methods.
    
\section{Theory}
\label{sec:theory}
%\subsection{semischematic model}

The semischematic model is defined by introducing in the ideal MCT 
equations for simple liquids an effective coupling parameter $\chi_R$ 
which models the caging effect of the molecular rotational motion on the
COM dynamics, i.e. the slowing down of the COM relaxation introduced by the 
angular degrees of freedom.
The resulting system of integro-differential equations\cite{ssm}
describes the time evolution of the normalized density-density COM 
correlation functions $\phi_q(t)=S_q(t)/S_q$,  $S_q(t)$ being 
the dynamic structure factor
\begin{equation}
S_q(t) = {1 \over N} < \rho_q(t)^* \rho_q(0)>
\end{equation}
and $S_q = S_q(0)$ the static structure factor.
The unknown $\chi_R$ is fixed once and for all by requiring that the
ideal glass transition temperature in the model $T_c^{MCT}$ coincides
with the temperature calculated form the analysis of experimental or
MD data.  
The imposed equality of the theoretical and MD temperatures is
very important because it allows to compare the numerical and the MD
data at the same reference $T$, i.e. with the same structure
factor and the same thermal broadening\cite{problemiconaltretecniche}. 
A larger value of $\chi_R$ 
corresponds to a stronger slowing down of the COM relaxation due to
the interaction with the rotational motion. If $\chi_R=1$ the semischematic model coincides with the standard MCT\cite{review-glass}.  

The physics described by the semischematic model is the same as for the ideal MCT. It predicts that
when $T < T_c^{MCT}$, $\phi_q(t)$ does not relax to 
zero at long times and the COM dynamics is frozen. Above $ T^{MCT}_c$
the correlator  $\phi_q(t)$  decays with a typical two step relaxation process,
characterized by the fast decay to a plateau value
(the non ergodicity parameter) followed by a slow relaxation to zero which
gets slower and slower as  $ T^{MCT}_c$ is approached ($\alpha$-relaxation).

The semischematic model describes 
the dynamic evolution of $\phi_q(t)$ in the
time region where the slow dynamics becomes dominant ($\alpha$-region) 
by the system of coupled equations
\begin{equation}
\phi_q(\hat{t})=m_q(\hat{t}) -
{d \over {d \hat{t}}} \int_0^{\hat{t}} ds \,\, m_q(\hat{t}-s) \phi_q(s)
\label{eq:dyn}
\end{equation}
where the time variable $\hat{t}$ is defined in terms of a 
characteristic time scale which diverges at $T^{MCT}_c$\cite{review-glass}.
The {\it memory function} $m_q$ is a quadratic functional of the 
correlations $\phi_q(t)$ themselves 
\begin{equation}
m_q [\phi_k(t)] = {\chi_R \over 2} \int{ {{d^3k} \over 
(2 \pi )^3} V(\vec{q},\vec{k}) \phi_k(t) \phi_{|\vec{q}-\vec{k}|}}(t) 
\label{eq:mq}
\end{equation}
and its increase on cooling is responsible for the 
slowing down of the relaxation process. The parameter $\chi_R$
enters in Eq. \ref{eq:mq} 
as a $q$-independent multiplicative factor, thus increasing the strength of the COM memory function.

The vertices in Eq. \ref{eq:mq}
are defined as
\begin{eqnarray}
V(\vec{q},\vec{k}) \equiv  S_q S_k S_{|\vec{q}-\vec{k}|} {1 \over {n q^4}}~~~~~~~~~~~~~~~~~~~~~~~~~~~~ \nonumber\\
\left[ {\vec{q}} \cdot 
\vec{k}  {{(1-S_k^{-1})} } +\vec{q} \cdot  
(\vec{q}-\vec{k})  { {(1-S_{|\vec{q}-\vec{k}|}^{-1} )} }
 \right]^2
\label{eq:v}
\end{eqnarray}
and are functions of the COM $S_q$ and inversely proportional to the density.

The infinite time limit $f_q$ of $\phi_q(t)$ can be calculated solving 
self--consistently on a discrete set of $q$ values 
the coupled integral equations

\begin{equation}
{f_q \over (1-f_q)}=m_q[f_k]
\label{eq:fq}
\end{equation}

using as input the number density $n$ and  $S_q$ 
which can be calculated from the MD data or measured experimentally.

Having chosen $\chi_R$ properly, at the ideal glass transition temperature
the solution $f_q$ of Eq. \ref{eq:fq}
jumps discontinuously from zero to a nonzero value, 
which defines the non ergodicity parameter $f_q^c$\cite{review-glass,gotze}. 
In what follows we will neglect the upper index in the non ergodicity 
parameter. 

In the supercooled liquid phase, after the
plateau $f_q$, $\phi_q(t)$ follows an initial power law decay (von Schweidler
law), ruled by a $q$ independent scaling exponent $b$, followed by a
stretched exponential relaxation (Kohlrausch-William-Watts law):
\begin{equation}
\phi_q(t) \sim A^K_q \exp \left[ {-{ \left( {t \over 
{\tau^K_q}} \right)^{\beta^K_q}}} \right] 
\label{eq:KWW}
\end{equation}
The range of validity of the von Schweidler law is strongly $q$ 
dependent\cite{franosch} and, therefore, it is worthwhile to consider also
the second order corrections:
\begin{equation}
\phi_q(t) \sim f_q-h_q^{(1)} \left({t \over \tau} \right)^b+h_q^{(2)}
\left({t \over \tau} \right)^{2b}+ O \left( (t/ \tau)^{3b} \right)
\label{eq:vs}
\end{equation}   
The exponent $b$ can be calculated solving
\begin{equation}
\lambda={{ \Gamma(1+b)^2 } \over { \Gamma(1+2b)}}
\label{eq:b-lambda}
\end{equation}
where $\Gamma$ is the Euler gamma function and the {\it exponent parameter} 
$\lambda$ is defined by
\begin{equation}
\lambda \equiv {1 \over 2}  \int\limits_0^\infty dq \int\limits_0^\infty  dk
\int\limits_0^\infty  dp \,\,
\hat{e}^c_q \,(1-f_k)^2 e^c_k {{\delta^2 m_q} \over {\delta f_k \delta f_p}}
 (1-f_p)^2 e^c_p
\label{eq:lambda}
\end{equation}
In Eq. \ref{eq:lambda}
$e^c$ and $\hat{e}^c$ are the right and left eigenvector corresponding to 
the  maximum eigenvalue of the stability matrix
\begin{equation}
C^c_{qk}[f_p] \equiv  {{\delta m_q[f_p]} \over {\delta f_k}}
 \left( 1-f_k \right)^2  
\end{equation}
evaluated at the critical point. The critical amplitude $h_q^{(1)}$ is
\begin{equation}
h_q^{(1)} = (1-f_q)^2 e^c_q
\end{equation}
while $h_q^{(2)}$ can be calculated as explained in Ref.\cite{franosch}.

In Sec.\ref{sec:results} we compare the predictions of the semischematic model
for the slow relaxation in a liquid of linear molecules with the
corresponding predictions of MMCT. To facilitate the reading of 
the present Article we present here a brief 
outline of the MMCT for linear molecules. We refer the 
interested reader to Refs. \cite{schilling,theis,schilling-vigo}.

According to MMCT the relevant variables for the description of the 
slow dynamics of supercooled molecular glass forming liquids 
are the generalized correlation functions
\begin{equation}
S_{l l'}^m(q,t) = {1 \over N} <\rho_{l m}(q,t)^* \rho_{l'm}(q,0)>
\label{eq:Slm}
\end{equation}
where the density functions are defined by
\begin{equation}
\rho_{l m}(\vec{q},t) = i^l \sqrt{4\pi} \sum_{j=1}^N e^{i \vec{q} \cdot \vec{x}_j(t) }
Y_{lm}(\Omega_j(t)) 
\label{eq:rholm}
\end{equation}
In Eq. \ref{eq:rholm} the sum runs over the $N$ molecules of the liquid,
$\vec{x}_j$ is the COM position of the $j$th molecule and 
$Y_{lm}(\Omega_j)$ are the spherical harmonics for
its Euler's angles. The index $l$ ranges over the whole set
of nonnegative integer numbers, while $m \in [-l,l]$.
For $l=l'=0$ the correlation function defined in Eq. \ref{eq:Slm} 
coincides with the COM correlator $S_q$ studied by the usual MCT and 
the semischematic model. 
Due to the lack of rotational invariance of the molecules the correlation 
functions depend both on modulus and orientation of the wave vector. In
Eq. \ref{eq:Slm} we have chosen a reference frame 
where $\vec{q}$ points in the direction of the $z$ axes, which offers
the advantage of diagonality of the correlators with respect to $m$.  
The evolution equations for the correlators in Eq. \ref{eq:Slm} 
are a generalization
of Eq. \ref{eq:dyn} for a single $q$ dependent correlator. The slowing
down of the relaxation is ruled by a infinite set of memory functions 
$[M_{l l'}^m(q,t)]^{\alpha \alpha'}$ which are quadratic functionals 
of the whole set of correlators (\ref{eq:Slm}). The extra indices 
$\alpha, \alpha' \in \{T,R\}$ are related to projection operations on the 
longitudinal translational $(T)$ and rotational $(R)$ currents.
For the aims of this Article we underline that in MMCT the time evolution 
of each correlator is coupled, through the memory functions, to every other 
correlator. This means that, due to the dependence of $[M_{0 0}^0(q,t)]^{T T}$
on both the translational and rotational correlators, the
dynamics of the COM correlator $S_{0 0}^0(q,t)$ is affected by the 
time evolution of every angular correlation function $S_{l,l'}^m(q,t)$ 
with $l \ne 0$.

With the cut-off $l \le 2$, the non ergodicity 
parameters $f_{l l}^m(q)= \lim_{t \to \infty} S_{l l}^m(q,t)/S_{l l}^m(q)$ 
have been
calculated for the liquid of LJ dumbbells in the approximation
of diagonality in $l$, i.e. $S_{l l'}^m(q,t) \approx  \delta_{l l'}
S_{l l}^m(q,t)$ and $[M_{l l'}^m(q,t)]^{\alpha \alpha'} \approx 
\delta_{l l'} [M_{l l}^m(q,t)]^{\alpha \alpha'}$\cite{schilling-vigo}.
We stress that this diagonality was also demanded for the static correlators 
$S_{l l'}^m(q)$, in order to keep the MMCT--equations as simple as possible.
The input of the calculations are the number density of molecules and all 
the static structure factors 
$S_{l l}^m(q)$ up to $l=2$ as evaluated from the set 
of MD data\cite{kob-dumb}.
In Fig. \ref{fig:nep} we show the results (from Ref. \cite{theis})
for the COM non ergodicity parameter
$f_{0 0}^0$ at two different temperatures (short and long dashed lines). 
The higher one, $T_t^{MMCT}=0.383$, 
corresponds to the MMCT theoretical freezing temperature for the translational
degrees of freedom, i.e. at $T_t^{MMCT}$ the COM $f_{0 0}^0$ jumps to a 
non zero value while the rotational dynamics is still in a liquid phase.
At the lower temperature $T_r^{MMCT}=0.310$ all the other non ergodicity
parameters $f_{l l}^m$ ($l \ne 0$) become nonzero, 
i.e. $T_r^{MMCT}$ is the MMCT theoretical transition 
temperature for the rotational dynamics. This splitting of critical 
temperatures, not observed in the simulation, is an artifact of
the diagonalization approximation and is not observed when 
the approximation $l \ne l'$ is waived\cite{private}.

\section{Model and Simulation Data}
\label{sec:model}

The model under investigation is an one--component system containing
$N = 500$ rigid diatomic molecules. Each molecule consists of two
atoms, labeled A and B, separated by a distance $d$. The
interaction between two molecules is build up by pair interactions
between the atoms, which are due to the LJ potentials
\begin{eqnarray}
V_{\alpha \beta}(r) = 4 \epsilon_{\alpha \beta} \left[ \left( \frac{
\sigma_{\alpha \beta}}{r} \right)^{12} - \left( \frac{\sigma_{\alpha
\beta}}{r} \right)^{6} \right] \nonumber
\\ \alpha,\beta \in \{A,B\}
\label{eq:potential}
\end{eqnarray}
with LJ parameters $\epsilon_{AA} = 
\epsilon_{AB} = 1.0, \epsilon_{BB} = 0.8$ and $\sigma_{AA} =
\sigma_{AB} = 1.0, \sigma_{BB} = 0.95$, i.e. $\epsilon_{AA}$ was
chosen as the unit of energy and temperature ($k_B = 1$) and
$\sigma_{AA}$ as the unit of length. The unit of time is then $\left(
(\sigma_{AA}^2 m)/(48 \epsilon_{AA}) \right)^{\frac{1}{2}}$ where $m$
is the mass of an atom which is chosen to be equal for both types of
atoms. The slight head--tail asymmetry of the dumbbell assures,
together with the choice of $d = 0.5$ as inter-atomic distance, a good
coupling between translational and rotational motion on the one hand
and avoids crystallization into a liquid crystalline phase and the
intersection of two dumbbells on the other. 

After equilibrating the system in the (N,p,T)--ensemble for times
which exceeded the relaxation times of the system even at the lowest
temperature, the production runs were carried out in the
microcanonical ensemble. To improve the statistics the data for each
temperature were averaged over at least 8 independent runs. Further
details about the simulation can be found in the references
\cite{kob-dumb} from which we take part of the data to be
compared with the theoretical results. 

The critical temperature for the COM which was determined from the
simulation by fitting the $\alpha$--relaxation--time and the diffusion
constant $D$ with power laws
\begin{equation}
\tau \propto (T-T_c)^{-\gamma} \qquad , \qquad D \propto
(T-T_c)^{\gamma}
\label{eq:powerlaws}
\end{equation}
according to MCT predictions, was found to be $T_c^{MD} = 0.477$.
The numerical values for the non ergodicity parameter $f_q$, 
the critical amplitude $h_q^{(1)}$ and the second order correction $h_q^{(2)}$
at the critical temperature $T^{MD} = 0.477$ are also taken from
reference\cite{kob-dumb}.
They were evaluated by fitting the decay from the plateau 
in $\phi_q(t)$ with the von
Schweidler law Eq. \ref{eq:vs}, including the second order correction.
From the same procedure the critical exponent $b = 0.55$ is also obtained
and, via Eq.\ref{eq:b-lambda},
one gets for the exponent parameter $\lambda$ the result $\lambda =
0.76$.
Furthermore we have examined the time dependence of the MD correlator
$\phi_q(t)$ in the $\alpha$--region by evaluating the amplitudes
$A_q^K$, stretching exponents $\beta_q^K$ and relaxation times
$\tau_q^K$ of a Kohlrausch--Williams--Watts fit (Eq. \ref{eq:KWW}).

\section{Results}
\label{sec:results}

\begin{figure}
\centerline{\psfig{figure=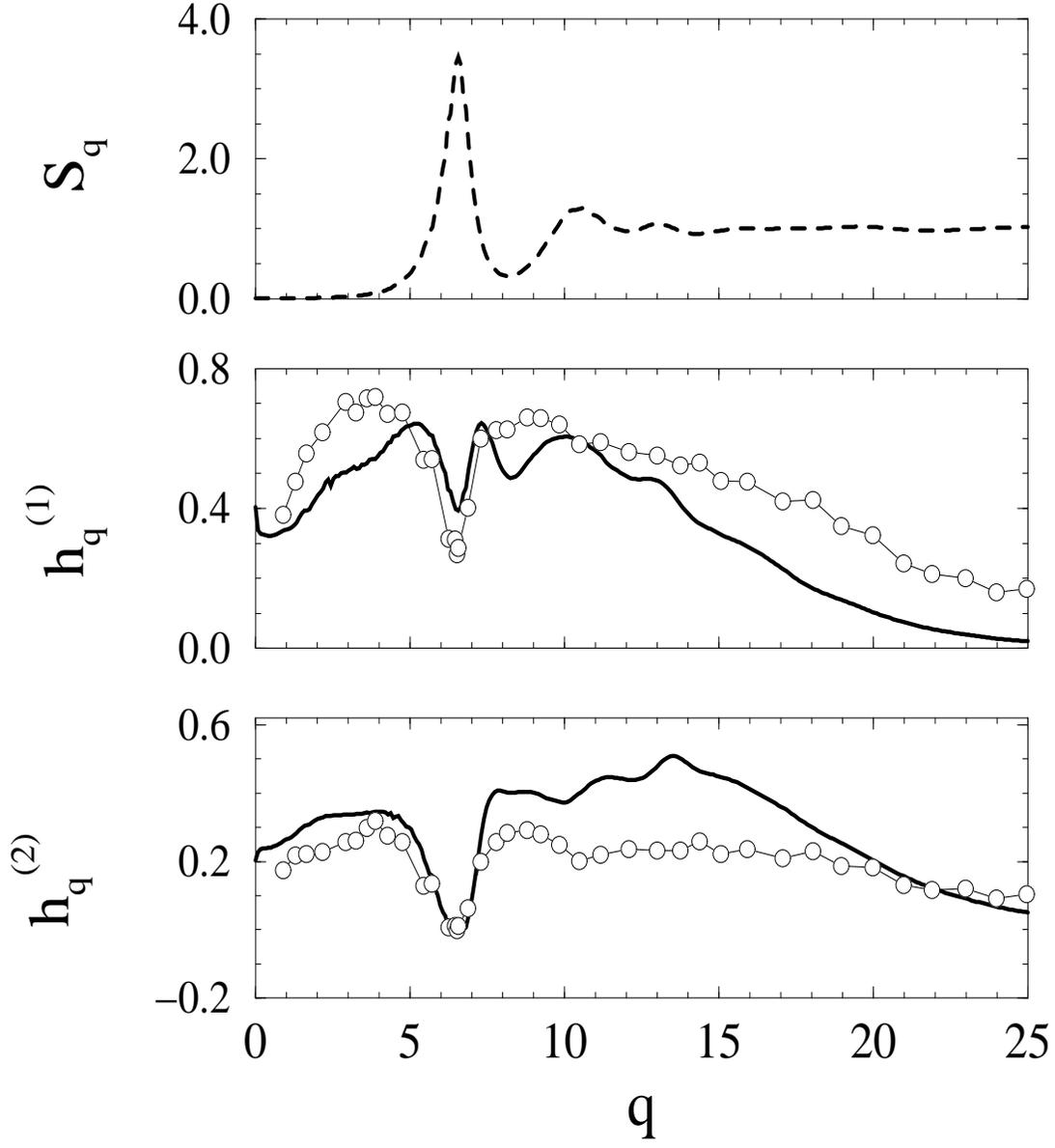,height=18cm,width=20cm,clip=,angle=0.}}
\caption{Critical amplitudes $h_q^{(1)}$ and $h_q^{(2)}$ in the 
von Schweidler law
Eq.\protect\ref{eq:vs} as predicted by the semischematic model (solid lines)
and as calculated from the MD data (symbols). The static structure factor $S_q$
used as input of the theoretical calculation is also shown.
The unit of $q$ is $2 \pi \sigma_{AA}^{-1}$.
} 
\label{fig:h1h2}
\end{figure}

\begin{figure}
\centerline{\psfig{figure=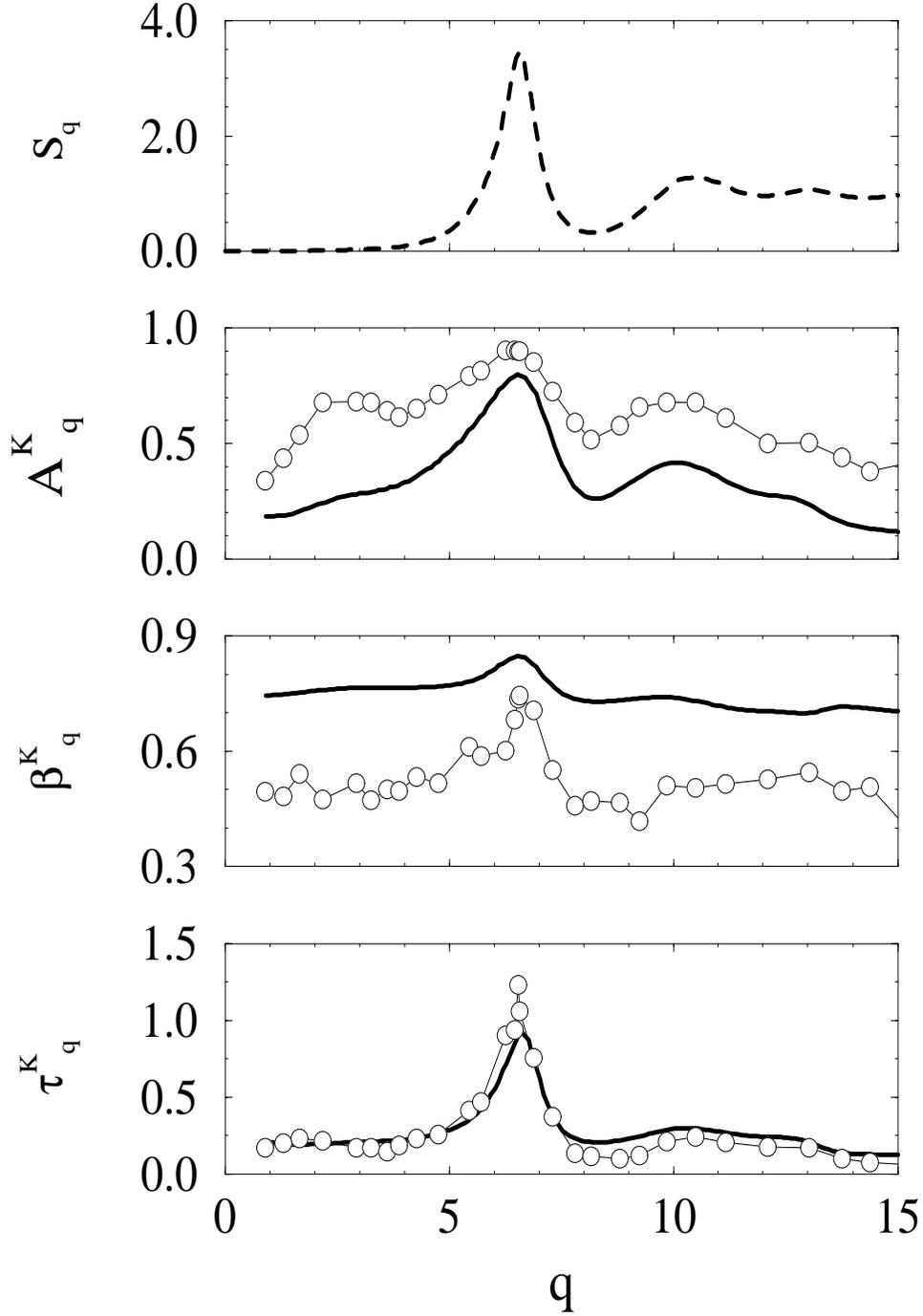,height=20cm,width=22cm,clip=,angle=0.}}
\caption{Amplitude $A^K_q$, stretching exponent $\beta^K_q$ and relaxation
time $\tau^K_q$ in the Kohlrausch-William-Watts law Eq. \protect\ref{eq:KWW}.
Lines are obtained fitting Eq. \protect\ref{eq:KWW} to the exact time dependent solution of 
Eq. \protect\ref{eq:dyn}  
while symbols are evaluated fitting the same law to the MD data.
The MD time unit for $\tau^K_q$ is  
$10^5((\sigma_{AA}^2m)/(48 \epsilon_{AA}))^{1/2}$ 
while the MCT relaxation times are arbitrarily scaled.
The unit of $q$ is $2\pi \sigma_{AA}^{-1}$. 
The static structure factor is shown as reference.
} 
\label{fig:dyn-dumb}
\end{figure}

We solve for the liquid of LJ dumbbells the semischematic
equations introduced in Sec. \ref{sec:theory} using as
inputs of the calculation the static structure factor $S_q$ as obtained
from the simulation (Figs. \ref{fig:h1h2},\ref{fig:dyn-dumb}) and the COM 
number density $n=0.719$ at $T^{MD} = 0.477$. 
By solving Eq.\ref{eq:fq} on a grid of 300
equispaced $q$ vectors, we find that the condition $T_c^{MCT}=
T^{MD}$ fixes the value of $\chi_R$ to $1.17$, which suggests that
the coupling between COM and angular degrees of freedom in LJ
dumbbells increases the COM memory function about $20\%$.  If compared
to the value $\chi_R =1.93$ found for SPC/E water\cite{ssm,sgtc}, this
result highlights the weaker hindering effect of the rotational motion
in a liquid of LJ linear molecules with respect to the
strong one observed in an hydrogen-bonded network-forming liquid. In
water the highly energetic hydrogen bonds build up the network
structure which is responsible for the caging of the molecules
in the glass phase. The motion of the COM of the molecules is completely
slaved to the breaking and reforming of the hydrogen bonds, i.e. it is
definitely correlated to the angular dynamics.

\begin{figure}
\centerline{\psfig{figure=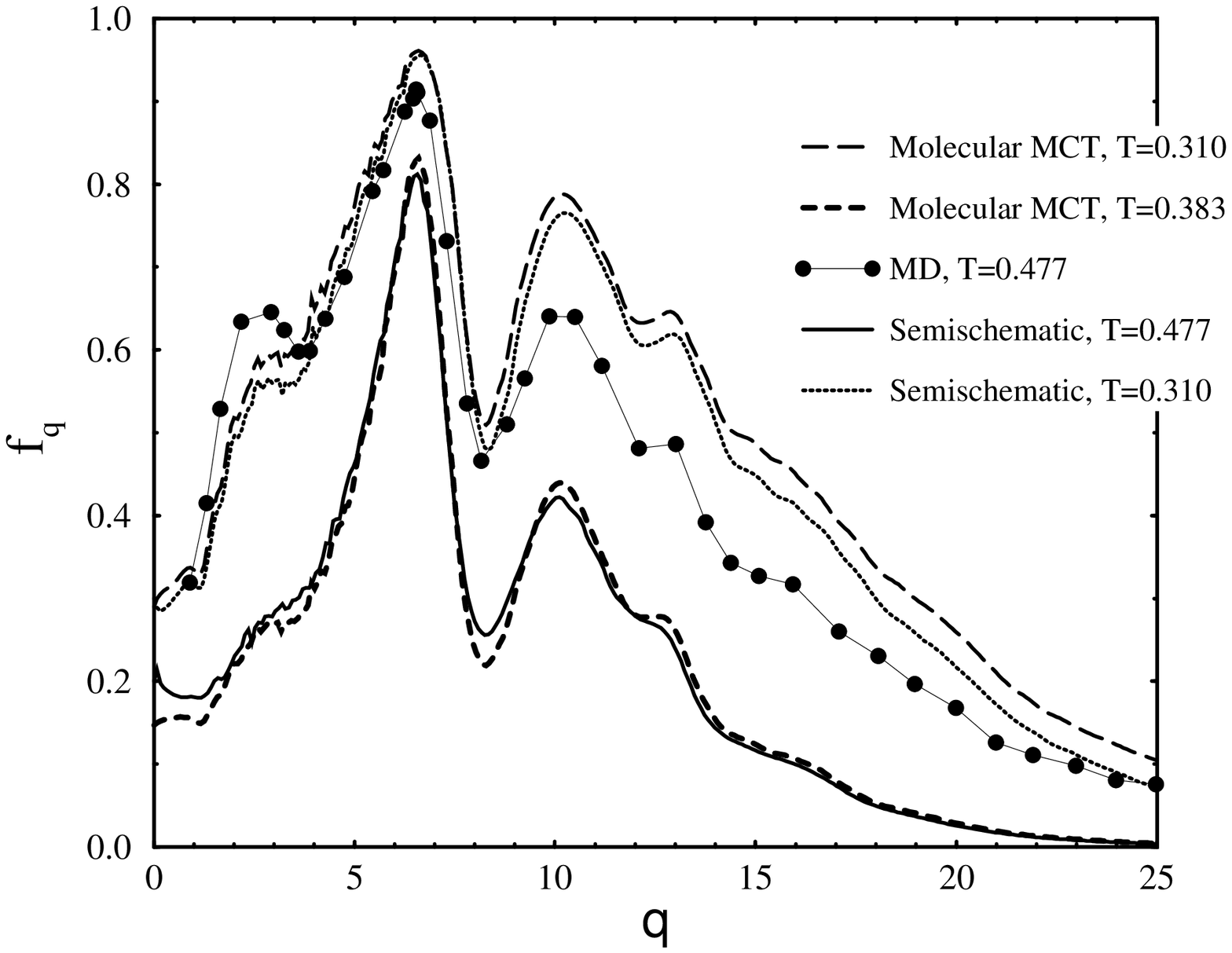,height=15cm,width=20cm,clip=,angle=0.}}
\caption{COM non ergodicity parameter $f_q$ as calculated by solving 
Eq. \protect\ref{eq:fq} (solid line) and as evaluated from the MD simulations
\protect\cite{kob-dumb} (symbols). For comparison also the COM non ergodicity 
parameters as predicted by MMCT are reported at two different temperatures
(dashed lines). The dotted line shows $f_q$ as predicted by the semischematic 
model at $T=0.310$. The unit of $q$ is $2 \pi \sigma_{AA}^{-1}$
while the unit of temperature is $\epsilon_{AA}$ (setting $K_B=1$).
The MMCT curves are from Ref. \protect\cite{theis}.     
} 
\label{fig:nep}
\end{figure}

The solution of  Eq.\ref{eq:fq} for $\chi_R=1.17$  as a function of the 
wave-vector $q$ is shown in Fig. \ref{fig:nep} together with 
the COM non ergodicity parameters as calculated
from the MD data. The theoretical $f_q$ oscillates in phase with
the MD data, but underestimates the amplitude, especially at large $q$ vectors.
The shoulder around $q=3$, which may be attributed to the 
rotational-translational coupling, as the orientational 
correlator $S_{11}^0(q)$ has a maximum at this $q$, is also underestimated.
This notwithstanding,  the semischematic $f_q$ 
captures the $q$-dependence of the 
non ergodicity parameter as calculated from the MD data.

The simplicity of the semischematic equations allows us to study, 
besides $f_q$, also the complete time relaxation of $\phi_q(t)$.
According to the theoretical predictions outlined in Sec. \ref{sec:theory}
we calculate the critical amplitudes $h_q^{(1)}$ and $h_q^{(2)}$ and 
the exponent $b$ which rule the early $\alpha$-relaxation behavior
(Eq. \ref{eq:vs}).

In Fig. \ref{fig:h1h2}, we present a comparison between the
theoretical predictions of the semischematic model for the critical
amplitudes $h_q^{(1)}$ and $h_q^{(2)}$ and the same
quantities as calculated by fitting Eq.\ref{eq:vs} to the 
density-density correlation
functions, i.e. with a quadratic fit in $t^b$. The
fitting coefficients $h_q^{(1)}/\tau^b$ and $h_q^{(2)}/\tau^{2b}$ can
be compared with the theoretical critical amplitudes after fixing, once
and for all, the q-independent time scale $\tau$, introduced by
the scale invariance of Eq. \ref{eq:dyn}.  The agreement is
satisfactory in a wide range of $q$ values.  The theoretical value for
the exponent parameter, as calculated from Eq. \ref{eq:lambda}, is
$\lambda=0.63$ to be compared with $\lambda=0.76$ as obtained by the
simulation, i.e.  the difference between the two is about $15\%$. This
yields a theoretical critical exponent $b=0.75$, while the exponent
calculated from MD data is $b=0.55$.  In the case of SPC/E water the
theoretical and MD values of $b$ coincide within the numeric error.
Such finding is consistent with the remarkable agreement of the
non ergodicity parameters over both the relevant peaks of the structure
factor.

The comparison between MD data and prediction of the 
semischematic model can be extended to the long time region. We solve the
complete dynamic set of equations (\ref{eq:dyn}) in the whole $q$ 
range\cite{algorithm}. 
We fit the stretched exponential law  (Eq. \ref{eq:KWW})
to the long time relaxation (late $\alpha$--region) and we compare 
the obtained amplitude $A^K_q$, relaxation time $\tau^K_q$
and stretching exponent $\beta^K_q$ with the 
corresponding quantities as calculated by fitting the MD relaxation. 
The comparison of the complete $q$ dependence of $A^K_q$, $\tau^K_q$ and
$\beta^K_q$ is shown in Fig. \ref{fig:dyn-dumb}. 
The theoretical and numerical relaxation times are in perfect agreement.
Less satisfactory, as in the case of water, is the theoretical prediction 
for $\beta^K_q$ for which the theory provides the correct qualitative 
$q$--dependence, but 
failing in amplitude up to $30\%$. The error in the values of the
stretching exponents is expected on the basis of the drastic
simplification adopted in the semischematic approach, which
condensate the coupling between the infinite set of angular correlator
and the COM correlator.  Indeed, in phenomena in which the decay of
correlation results from the sum of several independent relaxation
processes, a smaller $\beta$ indicates a wider distribution of
relaxation times\cite{st-relax}.
As expected $A^K_q$ has
the same behavior as $f_q$ both for theory and simulation.

The choice of the dumbbell liquid is particularly interesting 
because it allows a comparison between the theoretical predictions of the
semischematic model and those of MMCT which provides
a deeper understanding of the basic approximation in the model, 
i.e. the assumption that the coupling
between the COM and angular degrees of freedom can be quantified in a
multiplicative $q$--independent factor $\chi_R$. The comparison
requires a certain degree of care, because of the difference in the
theoretical estimate of $T^{MMCT}$, rather different from 
$T_c^{MD}$.  Indeed, while in the case of the semischematic model the
MD and the theoretical $f_q$ are evaluated at the same
temperature, in the case of MMCT, the non ergodicity parameters are
calculated using as inputs the structure factors from the simulations  
{\it but} evaluated at temperatures different from $T_c^{MD}$.
Fig. \ref{fig:nep} reports the COM MMCT non ergodicity parameters 
obtained in the approximations of diagonality and $l \le 2$ 
for the freezing of the COM dynamics only, and for the freezing of both 
COM and angular degrees of freedom. 

\begin{figure}
\centerline{\psfig{figure=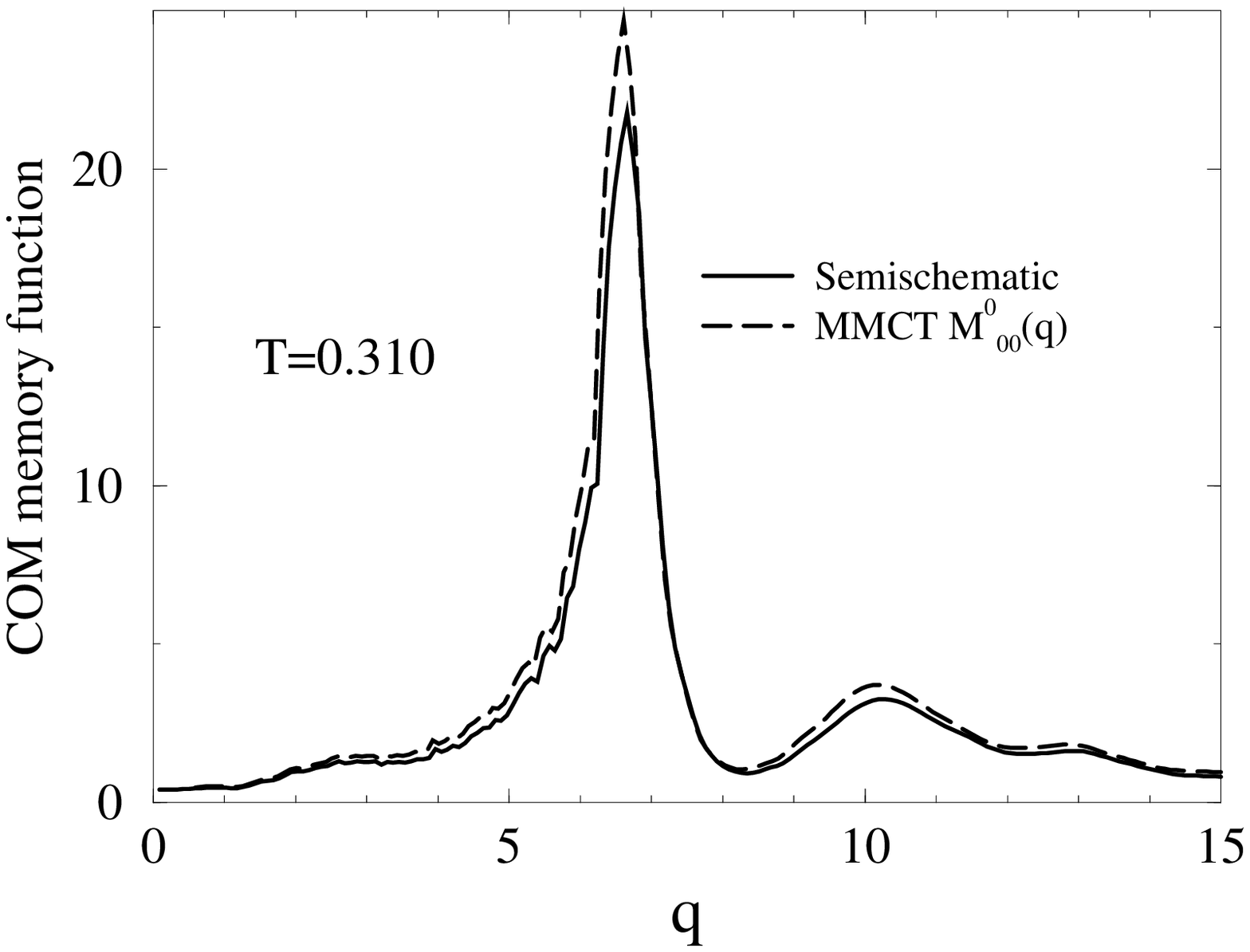,height=15cm,width=20cm,clip=,angle=0.}}
\caption{COM memory function as predicted by the semischematic model
Eq. \protect\ref{eq:mq} (solid line) and by MMCT (dashed line) at 
$T_r^{MMCT}=0.310$. The unit of $q$ is $2\pi \sigma_{AA}^{-1}$
while the unit of temperature is $\epsilon_{AA}$ (setting $K_B=1$).
} 
\label{fig:memory}
\end{figure}

A better insight into the comparison between the different theoretical 
approaches can be performed studying the non ergodicity parameter as 
predicted by the semischematic model at $T_r^{MMCT}$. Keeping fixed
once and for all the coupling $\chi_R=1.17$, we can solve Eq. \ref{eq:fq}
with the memory function Eq.\ref{eq:mq}. With this choice of temperature
the COM dynamics is in a deep glassy phase, both in the semischematic
and MMCT descriptions. Thus, being in a non ergodic phase, the COM dynamic 
structure factor has a finite long-time limit, which is shown in
Fig. \ref{fig:nep}. The semischematic (long dashed line) and MMCT
(dotted line) predictions are in perfect agreement. This
result can be enlightened by the comparison of the long time limits of
the semischematic memory function $m_q$ and the COM MMCT memory function 
$[M_{0 0}^0(q)]^{T T}$ at $T_r^{MMCT}$ (Fig. \ref{fig:memory}).
We recall that the angular correlators contribute to
$[M_{0 0}^0(q)]^{T T}$ as well as the COM correlator, while the semischematic
$m_q$ is ruled only by the COM $S_q$ and the value of $\chi_R$.   
Thus, the semischematic model gives rise to the complete
functional dependence of the COM memory function on the rotational 
relaxation in a very simple way, i.e. taking into account only
the functional dependence on the COM correlator and summarizing 
all the remaining coupling in the effective $\chi_R$. 
This means that the COM dynamics predicted by the semischematic model
almost coincides with the corresponding predictions of MMCT
in the diagonalization approximation. Furthermore the model has the 
advantage that the complete time evolution of the COM
correlator can be calculated with current computational resources.

\section{conclusions}

A theoretical (but relatively simple) description of the COM dynamics
in a molecular supercooled liquid is a relevant task which is
recently focusing a lot of efforts. The situation recalls 
the early day of the MCT for simple liquids,
when the dynamical equations were known but the exact numerical 
solutions were too difficult to handle. 
In that situation, simple approximations giving rise to a solvable 
set of equations were proposed and carefully studied.
This class of approximations, which is still 
extensively used  to interpret in a simple
way experimental results\cite{gotze-glicerolo,alba}, 
arises from the basic assumption that the $q$--dependence of 
$S_q$ can be reduced to a single representative $q_0$ vector, i.e.
$S_q \propto \delta (q-q_0)$\cite{schem}. 
In this approximation, which is named {\it schematic model},
the $q$--dependence is abandoned 
in favor of an exact description of one or two
representative correlators. 
In the same spirit, the {\it semischematic} model provides 
a method for studying the complete $q$--dependence of the 
COM dynamics, neglecting the angular degrees of freedom which are condensed 
in a single $q$-independent parameter $\chi_R$. 

Notwithstanding the drastic approximation intrinsic in the proposed
approach, we have shown in this Article that the model captures the
essential ingredients of the $q$ dependent static and dynamic
features of the COM $\alpha$-relaxation. 
The predicted non ergodicity parameter, relaxation time
and stretching exponent oscillate in phase, the critical amplitudes out of 
phase, with the COM structure factor, i.e. the model predicts the same
qualitative behavior observed in the simulation.  
The quantitative agreement with the MD data for $f_q$, $h^{(1)}_q$,  
$h^{(2)}_q$ and $\tau_q$ is satisfactory, especially for $q$ vectors 
close to the maximum of $S_q$, while the 
stretching exponent $\beta_{K}$ is overestimated.  
This reflects the major weakness of the
approach but at the same time clearly indicates the role played by the
angular degrees of freedom in controlling the dynamical evolution of
the center of mass.  If the present observations are discussed together
with the semischematic analysis of the dynamics of SPC/E water, a model which
mimics a liquid of strong directional hydrogen
bonds, it gets obvious that the value of $\chi_R$ 
is a measure for the strength of the roto-translational coupling. 

The comparison between the semischematic model and the MMCT 
approximate solutions for the non ergodicity parameter and the memory 
function also supports the validity of the assumption 
of a $q$--independent $\chi_R$.  Moreover, the coincidence of  
the theoretical critical $T$ and the MD one, 
opens the way to precise comparisons between theory and experiments.
Differently from the MMCT, the only required input in the model is the
COM $S_q$, a quantity which can be experimentally measured 
by suitable designed
neutron or x-rays scattering experiments\cite{toelle}.

\noindent
\section{Acknowledgment} 
R.S. and C.T. gratefully acknowledge
financial support by SFB--262.

\end{multicols}

\begin{references}

\bibitem{review-glass} W. G\"{o}tze, 
in {\it Liquids, Freezing and the Glass Transition}, 
Eds.:J. P. Hansen, D. Levesque and J. Zinn-Justin, Les Houches. 
Session LI, 1989, (North-Holland, Amsterdam, 1991).

\bibitem{gotze} W. G\"otze and L. Sj\"ogren, {\em Rep. Prog. Phys.} {\bf
55}, 241 (1992).

\bibitem{review} For recent reviews see for example 
C.A. Angell, Proc. 1996 Enrico Fermi Summer School in Physics, 
Italian Physical Society, in press. 
R. Schilling in {\it Disorder Effects on Relaxation Processes}, 
Eds. R. Richert and A. Blumen (Springer, Berlin 1994).

\bibitem{franosch-dumb} T. Franosch, M. Fuchs, W. G\"otze, M.R. Mayr and 
A.P. Singh, {\em Phys. Rev. E} {\bf 56} 5659 (1997).

\bibitem{schilling}
R. Schilling and T. Scheidsteger,{\em Phys. Rev. E} {\bf 56} 2932 (1997). 
T. Scheidsteger and R. Schilling, {\em Phil. Mag.} {\bf 76} 305 (1998). 

\bibitem{theis} 
C. Theis, Diplom Thesis Johannes Gutenberg Universit\"at (1997).

\bibitem{schilling-vigo} C. Theis and R. Schilling,
proceedings of the 3rd International Discussion Meeting on Relaxation in 
Complex Systems, Vigo (1997), to be published in {\em J. Non-Cryst. Solids}.

\bibitem{ssm} 
L. Fabbian, F. Sciortino, F. Thiery and P. Tartaglia, 
{\em Phys. Rev. E} {\bf 57} 1485 (1998).

\bibitem{kob-dumb} S. K\"ammerer, W. Kob and R. Schilling,
{\em Phys. Rev. E} {\bf 56} 5397 (1997).
S. K\"ammerer, W. Kob and R. Schilling, unpublished.

\bibitem{problemiconaltretecniche} 
Usually,
the predicted MCT critical temperature $T^{MCT}_c$ (or critical
density) is higher (lower) then the corresponding ``exact'' critical
temperature $T_c^{MD}$ calculated numerically via MonteCarlo or 
MD simulations.  Comparisons between theoretical predictions and
numerical results are affected by such inevitable difference in
temperature (or density). Indeed, the structure of the
liquid at which MCT would predict the dynamical transition may be
rather different from the structure of the liquid at $T_c^{MD}$.  The
phase relations between $q$ dependence of $S_q$ and the $q$ dependence
of $f_q$, $h^{(1)}_q$ and $h^{(2)}_q$ observed at $T_c^{MD}$ may be
different from the one at $T^{MCT}_c$. Moreover, the Lamb-Mossba\"uer
factor, defined as the non-ergodicity factor for the self density
correlation , is strongly $T$ dependent.  Thus extreme care is
requested on comparing MCT predictions and numerical findings if evaluated
at different $T$.

\bibitem{franosch} T. Franosch, M. Fuchs, W. G\"otze, M.R. Mayr and 
A.P. Singh, {\em Phys. Rev. E} {\bf 55}, 7153 (1997).


\bibitem{private} R. Schilling and C. Theis, private comunication.

\bibitem{sgtc} F. Sciortino, P. Gallo, P. Tartaglia, S.-H. Chen, {\em
Phys. Rev. E} {\bf 54} 6331 (1996). \\
F. Sciortino, L. Fabbian, S.H. Chen
and P. Tartaglia, {\em Phys. Rev. E} {\bf 56} 5397 (1997).

\bibitem{algorithm} We use the numerical algorithm
described in A.P. Singh, Ph.D. Thesis, Technischen 
Universit\"at M\"unchen, (1997). See also  
W. G\"otze {\em J. Stat. Phys.} {\bf 83} 1183 (1996).


\bibitem{st-relax} F. Sciortino and P. Tartaglia, 
{\em Physica A} {\bf 236}, 140 (1997). 


\bibitem{gotze-glicerolo} T. Franosch, W. G\"{o}tze, M.R. Mayr and A.P. Singh,
{\em Phys. Rev. E}  {\bf55}, 3183 (1997).


\bibitem{alba}
C. Alba-Simionesco and M. Krauzman, {\em J. Chem. Phys.},  {\bf 102} 
6574 (1995).


\bibitem{schem} 
E. Leutheusser, {\em Phys. Rev. A}, {\bf 29}, 2765 (1984). 


\bibitem{toelle} See for example
A. T\"olle, H. Schober, J. Wuttke and F. Fujara,
{\em Phys. Rev. E} {\bf 56} 809 (1997) and references therein.





\end{references}
\end{document}